# Water-mediated ordering of nanoparticles in electric field

Dusan Bratko[1,2], Christopher D. Daub[1] and Alenka Luzar[1]

[1]Department of Chemistry, Virginia Commonwealth University, Richmond, VA 23284-2006, aluzar@vcu.edu

[2]Department of Chemical Engineering, University of California, Berkeley, CA 94720-1462, dnb@berkeley.edu

**Abstract**

Interfacial polar molecules feature a strongly anisotropic response to applied electric field, favoring dipole orientations parallel to the interface. In water, in particular, this effect combines with generic orientational preferences induced by spatial asymmetry of water hydrogen bonding under confined geometry, which may give rise to a *Janus* interface. The two effects manifest themselves in considerable dependence of water polarization on both the field direction relative to the interface, *and* the polarity (sign) of the field. Using molecular simulations, we demonstrate strong field-induced orientational forces acting on apolar surfaces through water mediation. At a field strength comparable to electric fields around a DNA polyion, the torques we predict to act on an adjacent nanoparticle are sufficient to overcome thermal fluctuations. These torques can align a particle with surface as small as 1 nm$^2$. The mechanism can support electrically controlled ordering of suspended nanoparticles as a means of tuning their properties, and can find application in electro-nanomechanical devices.

## 1 Introduction

Electrowetting on macroscopic surfaces is usually characterized in terms of contact angle, $0 \leq \theta_c \leq 180°$, defined by Young's equation $\cos\theta_c = (\gamma_{sv}-\gamma_{sl})/\gamma_{lv}$, where $\gamma_{ab}$ is the surface free energy between solid (*s*), liquid (*l*) and vapor (*v*) phases[1,2]. A change from negative to positive $\cos\theta_c$ implies a transition from (partially) drying to wetting behavior. In the present context, when characterizing the interactions of solid surfaces with water, we refer to surfaces with $\theta_c > 90°$ as hydrophobic and those with $\theta_c < 90°$, hydrophilic. Discussion of alternative conventions can be found in Refs. [3,4]. Surface wettability can be enhanced by application of electric voltage driving dipolar water molecules to the field-exposed region.

The macroscopic relation due to Young and Lippmann[5] can describe electrocapillarity in a planar confinement:

$$\cos\theta_c = \frac{\gamma_{sv} - \gamma_{sl}}{\gamma_{lv}} - \frac{W_{el}(V)}{2\gamma_{lv}} = \cos\theta_c^o - \frac{W_{el}(V)}{2\gamma_{lv}} \qquad (1)$$

Here $W_{el}(V)$ is the change in electrostatic free energy per unit area, associated with surface spreading of the liquid, wetting both walls (hence the factor 1/2), $V$ is the voltage across the interface, and $\theta_c^o$ the contact angle in the absence of electric field. The form of $W_{el}$ depends on system geometry and material properties but is generally presumed to be proportional to the areal electric capacitance of the interface, $c$, and the potential drop across the interface squared, $W_{el} \sim -\frac{c}{2}V^2$.[1,2]

Molecular level studies of these phenomena[6-9], however, can bring in an entirely new perspective. In our recent work we demonstrated the importance of molecular scale events in understanding and making quantitative predictions of electrowetting phenomena in a nanoscale system.[8,9] In particular, we highlighted the coupling between interfacial hydrogen bonds and molecular alignment in electric field. This coupling modulates the dependence on field direction and polarity in contrast with conventional picture in a macroscopic system discussed above; system size plays a crucial role.

These findings are based on our molecular simulations performed in two very different scenarios: a sessile nanodrop on an atomic graphene surface[8] and water in a nanopore[9]. The droplet scenario[8] is a miniature of the experimental setup of a drop inside a capacitor described by Bateni *et al.*[10,11] Except for using a uniform capacitor field, it resembles the case of a dielectric drop atop an electrode[1]. Here, surface tensions change since the field tends to be stronger at droplet interfaces. In the second scenario[9], water from a field-free bulk phase is driven into a planar hydrophobic confinement exposed to electric field. In both scenarios, we demonstrated a strong dependence of wetting on the *direction* of applied electric field, distinct from the macroscopic picture where the effect depends only on field squared (Eq. 1). In a nanoscale confinement water wets a surface better when the applied electric field is parallel than when it is perpendicular to the interface.

In perpendicular field, polarity can also make a difference. Even in the absence of directional forces between the walls and liquid molecules, wetting by water is asymmetric, enhanced at the wall at which the field points into the aqueous phase and almost unchanged at the opposite side[9] (*Janus* interface)[12]. We explain the sign-sensitive response to the field[9] in terms of orientation preferences of surface molecules.[13-16] We predicted such preferences, associated with water hydrogen bonding, long ago from mean-field arguments[13,15] and they were confirmed in simulation[14] and experiment[16]. The observed anisotropy in electrowetting is a new nanoscale phenomenon that has so far been elusive as, in current experimental setups, surface molecules represent a very low fraction of all molecules[10] affected by the field. This is different from spectroscopic techniques where molecules will sense the confinement effect without necessarily residing in the solid/liquid contact layer[17,18].

Field direction and polarity are therefore important determinants of electro-wetting effects in a nanoporous material. Conversely, for fixed field direction and freely rotating nanotubes (pores), or nanoparticles, surface energetics will favor the alignment of the particle surface with electric field suggesting a novel mechanism to address structural order in a nanomaterial through controlled molecular or supramolecular charge distribution. We explore such a scenario in this work.

We consider situations where the external field bears no direct influence on nonpolar nanopore walls; hence the presence of water is essential. The torque $\tau$ on a wall of area $A$ is $\tau(E,\phi) = -A\, \partial \gamma_{sl}(E,\phi)/\partial \phi$, where $\phi$ is the angle of the field direction relative to the wall. Free energy differences between aligned and perpendicular nanopores we showed in our earlier work[9] indicate the effect may suffice to align a water-filled nanotube in a field $E$ of strength comparable to that in the double layer next to a DNA polyion[19], synthetic polyelectrolyte[20], or ionic-surfactant assembly[21,22], $E\varepsilon_r \equiv E_0 \sim 1$ V Å$^{-1}$ ($\varepsilon_r$ is the relative permittivity).

We use molecular simulations to assess the magnitude of this solvent mediated effect. Specifically, we estimate the strength of orientational forces water can mediate between external field and nanopores. For a range of imposed field directions and strengths up to those next to ionized polymers[19], ion channels[23,24], and electrochemical electrodes[25], we determine forces and torques on water-filled nanopores with parallel walls for different wall-water affinities (measured in terms of varying water contact angle) and at several distances apart. For the above field strengths, we show that the work necessary to rotate the nanopore by an angle $\delta\phi$ away from full alignment, $w(E,\phi) = \int_0^{\delta\phi} \tau(E,\phi)d\phi$, exceeds thermal energy, $k_B T$ for wall areas as small as ~ 1 nm$^2$ and angular displacement ~ $\delta\phi$ $O(10^o)$.

The planar pore models we employ are defined in the next section. We then present results illustrating the role of field direction on asymmetric water structure, hydrogen bond populations, and water polarization in a field-exposed nanopore. We show how these effects relate to free-energy dependence on the angle between the applied field and pore walls, leading to a torque that drives the pore into alignment with the field. Our principal results demonstrate a notable water-mediated force toward interface alignment with the field. We show that only contact-layer water is engaged in this mediation, however, the attraction between water molecules and confinement walls bears only a minor influence. While we exemplify the aligning mechanism in the case of nanosized planar confinements, the effect is clearly more general and will operate at all smooth aqueous interfaces. Presently, we do not consider systems with specific directional forces between wall molecules and water but it is clear introduction of such forces would result in additional orientational trends superimposed on the more general effect addressed in the present study.

## 2 Dependence of free energy on field direction

The driving force toward interface alignment with the applied field $E_0$ can be quantified in terms of the change in wall/water surface tension as a function of the angle $\phi$ between field direction and the interface. When water in a planar, field-exposed confinement maintains equilibrium with a field-free bulk phase, the pertinent work function is the wetting surface free energy, $\sigma$, here defined as

$$\sigma = \left(\frac{\partial \Omega}{\partial A}\right)_{\mu,D,T} \quad (1)$$

where $\Omega$ is the grand potential and $A$ the water/wall contact area (the wetted area of confinement walls). Further, $\mu$ is chemical potential, $T$ temperature and $D$ the width of the confinement. For constant $D$, area $A$ is related to volume $V$ of confined water, $A=2V/D$, leading to an alternative expression for $\sigma$:

$$\sigma = \frac{D}{2}\left(\frac{\partial \Omega}{\partial V}\right)_{\mu,D,T} = -\frac{D}{2} P_{\parallel} \qquad (2)$$

where $P_{\parallel}=(P_{xx}+P_{yy})/2$ represents the average of the lateral components of the pressure in the confinement. In the absence of the field ($P_{\parallel}=P_{xx}=P_{yy}$), the quantity $\sigma$ can be identified with the surface tension difference $\Delta\gamma=\gamma_{sl}-\gamma_{sv}$. As we have shown in the preceding work[9], when confined water is polarized by applied electric field, the wetting surface free energy, $\sigma$, depends on the angle $\phi$ measuring the direction of the field relative to confinement surfaces. In general, surfaces aligned with the field ($\phi=0$) appear more hydrophilic than the ones perpendicular ($\phi=\pi/2$) to the applied field. In the present study, we determined the dependence of $\sigma$ on $\phi$ for a set of surfaces with different zero-field contact angles $\theta_c \sim 135°$, $93°$ and $69°$. In view of the strong dependence of pressure and $\sigma$ on the liquid density in the pore, itself a function of $\phi$, these calculations require open ensemble ($\mu VT$) simulations. Knowing $\sigma$ as a function of angle $\phi$ in an open ($\mu VT$) system also provides information about the torque $\tau$ acting on non-aligned confinement surfaces at specified field strength $E_0$ and the angle between the surface and the field, $\phi$:

$$\frac{\tau}{A} = \left(\frac{\partial \sigma}{\partial \phi}\right)_{\mu,V,T} = \frac{1}{A}\left(\frac{\partial \Omega}{\partial \phi}\right)_{\mu,V,T} \qquad (3)$$

If we are interested only in the torque exerted by the field, and not in the absolute values of $\sigma$, direct calculation of $\tau$ is also possible. Beginning with the relation

$$\Omega(\mu,V,T) = -k_B T \ln \Xi \qquad (4)$$

where $\Xi = \sum_N \sum_{\text{states } i} e^{-\frac{U_i}{k_B T}+\frac{\mu N}{k_B T}}$ is the grand canonical partition function,

we can express the torque (per unit area $A$) as

$$\frac{\tau}{A} = -\frac{1}{A}\frac{\partial \Omega}{\partial \phi} = \frac{k_B T}{A}\frac{\partial}{\partial \phi}\ln\Xi = \frac{k_B T}{A}\frac{\sum_N \sum_{\text{states } i} -\frac{1}{k_B T}\frac{\partial U_i}{\partial \phi} e^{-\frac{U_i}{k_B T} + \frac{\mu N}{k_B T}}}{\Xi} = -\frac{1}{A}\left\langle \frac{\partial U}{\partial \phi}\right\rangle_{\mu,V,T} \quad (5)$$

Note the difference between the grand potential derivative $\langle \partial U / \partial \phi \rangle_{\mu VT}$ in Eq. 5 and the energy derivative $(\partial \langle U \rangle / \partial \phi)_{\mu VT}$. An analogous relation for fixed $NVT$ obtains the torque in the canonical ensemble. Our simulation results (Section 3) confirm equality of torques determined as derivatives of $\sigma$ (Eq. 2) with those from direct calculation, Eq. (5).

## 3 Models and methods

### 3.1 Model systems

The model system consists of a confined water slab between parallel insulating walls whose uniformly distributed sites interact with water molecules through Lennard-Jones potential but carry no charges or dipoles. The thickness of the walls exceeds the range of water-water and water-wall interacting site pair potentials. To describe the interactions between water molecules we apply the SPC/E[26] potential, which has been commonly used in related studies[6-9, 25, 27]. The presence of external field may suggest the use of polarizable models, however, even for the non-screened fields $E_0$, the strengths we consider are weak compared to molecular and ionic fields that lead to significant polarization of a water molecule. According to fig. 6 of ref.[28], for fluctuating-charge (TIP4P-FQ)[29] polarizable force field of water, the strongest field considered in the present study, $E_0$=0.4 VÅ$^{-1}$, will produce about 1% change in the average dipole moment of a water molecule. Further, recent comparisons between different interaction-site simulation models of water show that calculated surface tensions of water are relatively robust with respect to details of the model potentials[30].

Interaction between water molecules and smooth confinement walls is described by the integrated (9-3) Lennard-Jones potential[9, 14, 31-33]

$$u_w(z) = A\left(\frac{\sigma_{wO}}{D' \pm z}\right)^9 - B\left(\frac{\sigma_{wO}}{D' \pm z}\right)^3 \quad . \quad (6)$$

Here, $A = 4\pi\rho_w \varepsilon_{wO} \sigma_{wO}^3 / 45$ and $B = 15A/2$[33], $D' = D/2$, $z$ is the distance of water oxygen from the slit midplane with walls at $z = \pm D'$, $\rho_w$ is the presumed uniform number density of interacting sites of wall material, $\varepsilon_{wO}$ and $\sigma_{wO}$ are Lennard-Jones potential depth and size parameters for water oxygen-wall site pair, and the sum or difference is used in the denominators to describe interactions with a wall at –D' or D', respectively. $\varepsilon_{wO}$ and $\sigma_{wO}$ are Berthelot's means for water oxygen and wall site Lennard-Jones parameters given in Table I. According to the above definitions, the slit width actually occupied by water molecules will be close to $D-\sigma_{wO}$.

To describe hydrophobic, paraffin-like walls, $w(1)$, we used Lennard-Jones parameters from Lee et al.[14] Walls with increased hydrophilicities, $w(2)$ and $w(3)$, were modeled following the approach of Jaffe and coworkers[34], who calibrated the strength of water-wall Lennard-Jones interaction to match the desired water contact angle on the specified surface. Using wall site densities, $\rho_w$, and sizes, $\sigma_w$, from the hydrocarbon wall model[33], and values for $\varepsilon_{w(1)O}$ and $\varepsilon_{w(2)O}$, interpolated from Table IV of Ref.[34] we obtain the following wall/water contact angles: $\theta_c(1) = 135°$, $\theta_c(2) = 93°$, and $\theta_c(3) = 69°$, all with error bars of $\pm 3°$. Contact angles were estimated by Young's equation and respective interfacial tension differences, $(\gamma_{sv}-\gamma_{sl})$, along with the surface tension for SPC/E water calculated at identical conditions in our earlier work, described in the supplement to Ref.[9]

**Table I.** Lennard-Jones parameters of different species, $\alpha$ (oxygen and hydrogen atoms of water and interacting sites of the walls, $w(i)$).

| $\alpha$ | $\varepsilon_\alpha$/kJ mol$^{-1}$ | $q_\alpha/e_o$ | $\sigma_\alpha$/Å |
|---|---|---|---|
| O | 0.6502 | -0.8476 | 3.1656 |
| H | - | 0.4238 | - |
| $w(1)$ | 0.6483 | - | 3.754 |
| $w(2)$ | 3.45 | - | 3.754 |
| $w(3)$ | 5.00 | - | 3.754 |

**Table I.** Lennard-Jones parameters of different atomic species, $\alpha$ (oxygen and hydrogen atoms of water and interacting sites of the walls, $w(i)$).

We consider two gap widths $D = 16.4$ Å and 27 Å. In the hydrophobic confinement, $w(1)$, the lower of the two widths is just above the threshold wall separation of mechanical instability (spinodal) for capillary evaporation[4, 33, 35]. We explore the effect of the field on the water phase behavior in a separate study.[36] Finite size effects were mitigated by the use of periodic boundary conditions along lateral directions; a 1 nm spherical truncation of intermolecular potentials was implemented following previous studies[9, 31]. Temperature was held at 298 K in all calculations.

We maintain equilibrium with ambient aqueous environment by using open ensemble Monte Carlo simulations (*GCMC*) following the procedures we described earlier.[9, 32, 33, 37, 38] Excess chemical potential for water $\mu_{ex}$=-12.1 $k_BT$ was selected to secure vanishing pressure in bulk water in the absence of applied field. Wetting free energies, $\sigma$, are calculated from average lateral components of pressure tensor, $P_\parallel$, Eq. (2).

For the analysis of water hydrogen bonding we use our usual definition based on geometric criteria[39] with the cutoff value $R_{OH}^c$ = 2.45 Å, the OH intermolecular distance dictated by the 1$^{st}$ minimum in the corresponding radial distribution function for SPC/E water, and a maximum angle between O-O vector and the covalent O-H bond of $\alpha^c$ = 30$^0$. This is the angle at which the average number of hydrogen bonds per water molecule is within 10% of the asymptotic value for large $\alpha^c$. Interestingly, this threshold angle does not change up to applied field $E_0$ of 1.5 V/A.[36]

### 3.2 Pressure tensor calculation

Average pressure tensor components $P_\perp = P_{zz}$ and $P_\parallel = (P_{xx} + P_{yy})/2$ are calculated from energy differences $\Delta U_\alpha$ associated with uniform scaling of molecular coordinates $\alpha$ ($\alpha$ = z or x,y) and volume change $\Delta V_\alpha$. Here, $\Delta U_\alpha$ comprises changes in intermolecular and water-wall interactions. Scaling of molecular center-of-mass positions has no effect on interactions with the applied electric field.

$$P_{\alpha\alpha} = \rho kT + \lim_{\Delta V_\alpha \to 0} \frac{kT \ln <exp(-\frac{\Delta U_\alpha}{kT})>}{\Delta V_\alpha} \qquad (7)$$

Forward and backward scaling was employed for improved accuracy. A coordinate scaling factor $f = 1 \pm \delta$ with $\delta = 10^{-5}$ was chosen empirically for optimal compromise between round-off errors (decreasing with increasing $\delta$) and the numerical accuracy of the finite difference approximation employed in Eq. 7 (improved upon decreasing $\delta$). Variation of $\delta$ within the interval $10^{-6} \leq \delta \leq 10^{-4}$ revealed no appreciable effect on calculated pressure components The calculated normal component of the pressure tensor, $P_\perp$, agreed within numerical uncertainty with wall pressure calculated directly from wall/water forces as described in our earlier work[33].

## 4 Results and Discussion

We first present our results for water density profiles in a field-exposed slit for different angles $\phi$ of incoming field of strength $E_0 = 0.2$ VÅ$^{-1}$. The slit width is 1.64 nm and the walls are composed of hydrophobic material with water contact angle 135°, model $w(1)$ in Table I. As shown in Fig. 1a, water density in the interior of the slit oscillates around that of the field-free bulk phase equilibrated with the confinement but is elevated in the immediate vicinity of the walls. The density increase depends on the angle $\phi$, being most pronounced when the field is parallel to the walls ($\phi=0$). In perpendicular field, electrostriction is weaker overall and strong asymmetry in the density profile is observed[9]. Similar asymmetry has been observed by other groups[6, 7, 25]. We show that density remains almost unaffected by the field on the r.h.s. wall where molecular alignment with the field is least favorable for hydrogen bonding. The curve corresponding to $\phi=\pi/4$ interpolates between the two extreme cases corresponding to $\phi=0$ and $\pi/2$.

Field-induced changes in hydrogen bond populations are shown in Fig. 1b. We demonstrate stronger wetting to coincide with higher population of hydrogen bonds. Interestingly, the number of hydrogen bonds per molecule $<n_{HB}(z)>$ appears to increase in the field in all cases shown. A slight reduction is observed only when the perpendicular field tends to orient both water hydrogens toward the r.h.s. wall, an orientation allowing at most two hydrogen bonds per molecule. In examples shown in Figures 1a-1d, the field is either parallel to confinement walls ($E_z = 0$) or contains a positive component perpendicular to walls ($E_z > 0$), i. e. pointing from left to right. Nearly unperturbed hydrogen bonding at the negative $z$ wall with the field at an angle $\phi=\pi/4$ reflects the flexibility of the hydrogen-bond network. Water molecules can keep hydrogen bonds while accommodating small perturbations from the optimal molecular alignment with the walls, but resist following the field when the deviation approaches $\pi/2$. Because of angle preferences imposed by hydrogen bonds, in perpendicular field, polarization of water is stronger at the left wall where the fields directs water hydrogens away from the wall (Fig. 1c). When the field is parallel to the walls, on the other hand, the tendency to optimize hydrogen bonding cooperates with already existing trend toward near-parallel alignment[14, 16, 31].

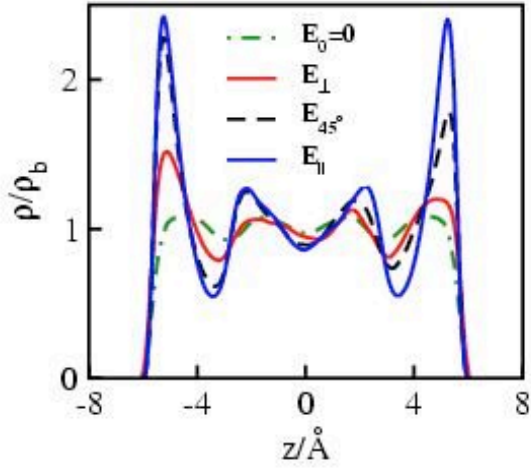

**Fig. 1a** Water density profiles in a 16.4 Å wide slit subject to electric field $E_0 = 0.2$ VÅ$^{-1}$ for different angles ϕ between the walls and incoming field. The green curve shows the density profile at $E_0=0$.

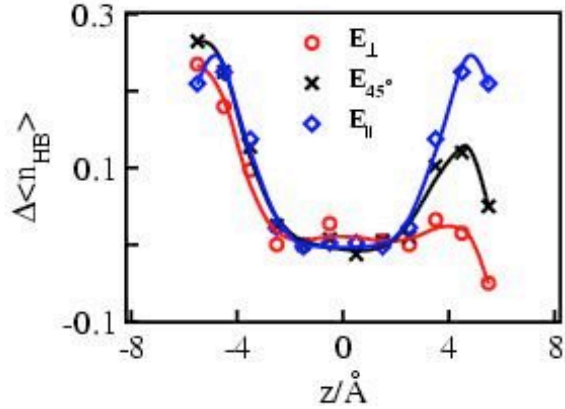

**Fig. 1b** The change in the average number of hydrogen bonds per molecule due to the applied field $E_0 = 0.2$ VÅ$^{-1}$ at different field angles ϕ relative to the walls.

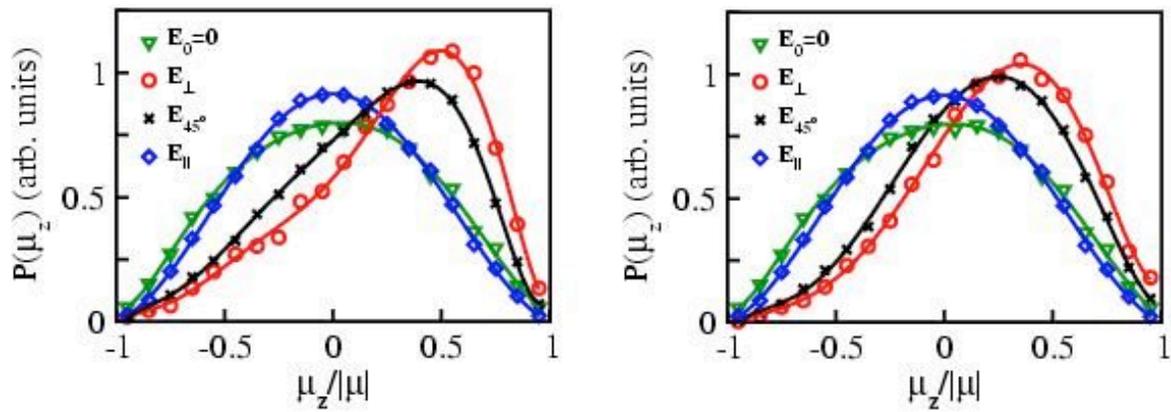

**Fig. 1c** Average projections of molecular dipole normal to the wall in the absence of applied field and in field of strength $E_0 = 0.2$ VÅ$^{-1}$ for different field/wall angles ϕ. Comparison between l.h.s. and r.h.s. walls

shows a stronger alignment with normal field on the left wall where the field orients water hydrogens into the liquid.

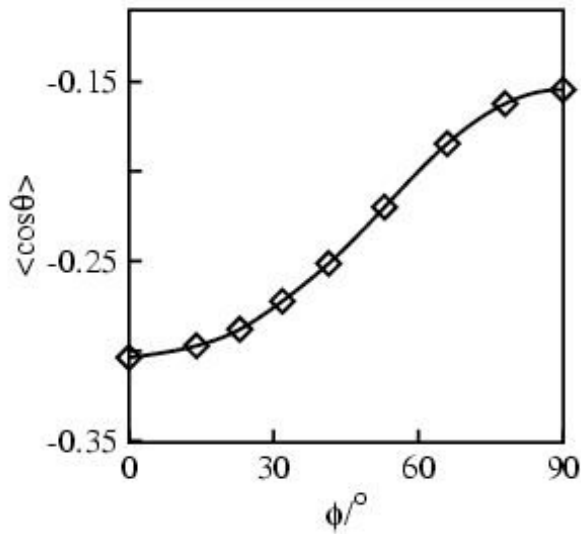

**Fig. 1d** Extent of dipole alignment with the field, measured in terms of $<\cos\theta> = <E_0\mu>/|E_0\mu|$, as a function of the angle between the direction of incoming field and confinement walls, $\phi$. Confinement width $D = 16.4$ Å and the water-wall contact angle is 135°.

Consistent with easier water polarization in the direction along the surfaces, Fig. 1d, electrostriction is strongest when the field is parallel to confinement walls. An interesting behavior is revealed when monitoring average numbers of water molecules, $<N>$, accommodated in slits with varied wall contact angles, Fig. 2. A comparison between wall types $w(1)$, $w(2)$, and $w(3)$ (see Table I) with contact angles $\theta(1) = 135°$, $\theta(2) = 93°$, and $\theta(3) = 69°$ shows expected differences in $<N>$ in the absence of the field. Application of perpendicular field results in considerable electrostriction in the hydrophobic pore characterized by high compressibility. This compressibility is associated with the existence of depleted water layers next to strongly hydrophobic plates ($\theta = 135°$)[40, 41]. In the field, available space is gradually used up to accommodate additional molecules attracted into the system. Pores with contact angles $\theta = 93°$ or 69°, on the other hand, are filled more tightly even in the absence of the field. Low compressibility of the liquid next to hydrophilic walls reflects weak density fluctuations comparable to those in the bulk phase. A similar regime is approached in hydrophobic pores after we suppress fluctuations by applying an external electric field to the system.

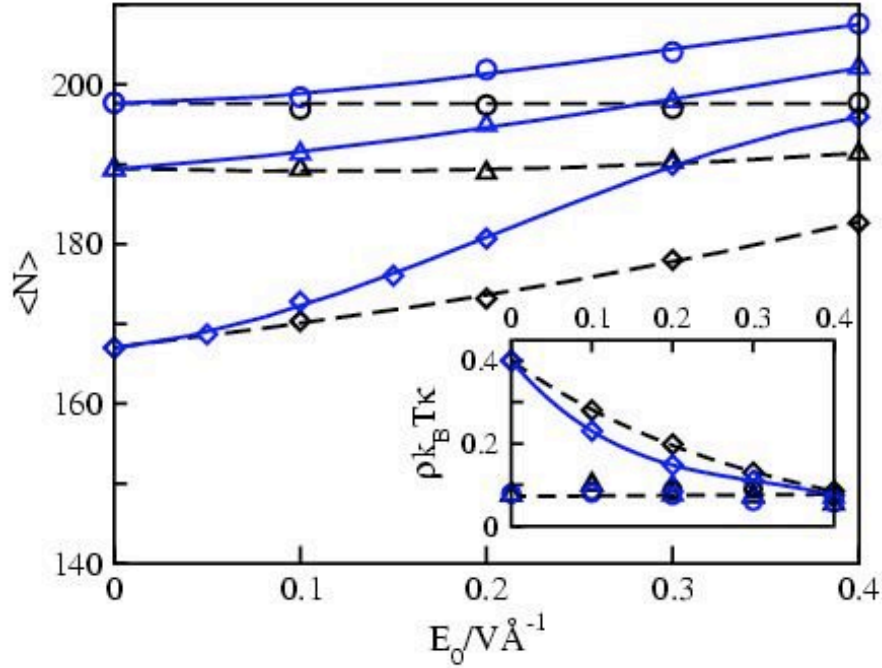

**Fig. 2** Average number of water molecules $<N>$, and reduced compressibilities $\rho \kappa k_B T = (<N^2> - <N>^2)/N$ in an open slit with periodic box dimensions $21 \times 21 \times 16.4$ Å$^3$ as a function of field strength $E_0$ for wall-water contact angles $\theta_c = 135°$ (diamonds), $93°$ (triangles) or $69°$ (circles) under perpendicular (dashed) or parallel fields (solid lines).

In view of the low initial compressibilities in pores with lower contact angles, only a slight density increase is observed in the perpendicular field. In a parallel field, electrostriction is noticeable even in the hydrophilic pore, $\theta = 69°$, again reflecting stronger affinity for water when the field is applied laterally along the walls.

Enhanced solvation of walls aligned with the field suggests considerable work is needed to rotate the plates away from a field-aligned orientation. Wetting surface free energies, $\sigma$, therefore depend on the angle $\phi$ between the field and the walls. This dependence is illustrated in Figs. 3a and 3b for all three wall types at two pore widths $D = 16.4$ and $27$ Å. For narrow hydrophobic pores, it has been shown that electric field reduces the threshold wall separation for capillary evaporation[42]. Our results demonstrate the effect will also depend on field direction, being maximal when the field is parallel to the walls, as is the case in ion channels.

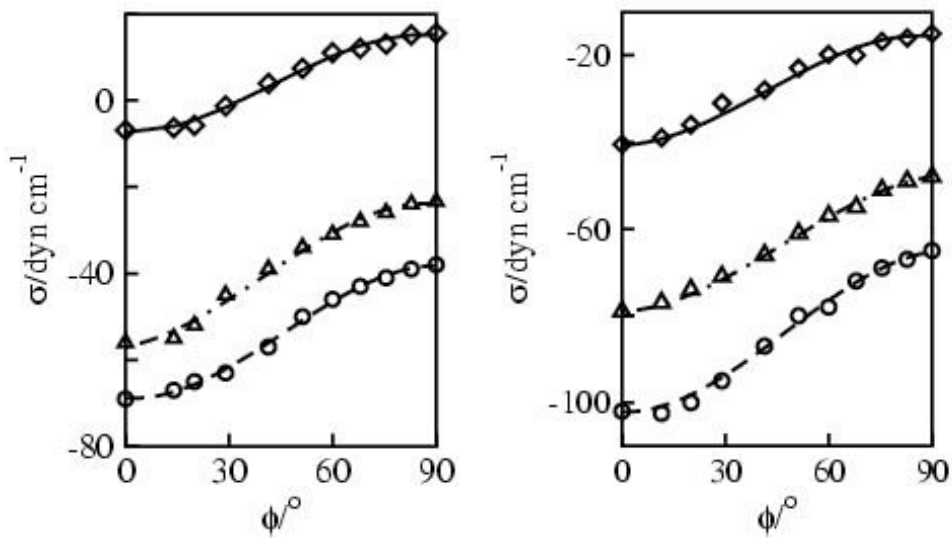

**Fig. 3** Wetting free energies of slits of width $D$ = 16.4 (left) or 27 Å (right) in electric field of strength $E_0$=0.2 VÅ$^{-1}$ as functions of field-wall angle $\phi$. Wall-water contact angles are 135° (diamonds), 93° (triangles) or 69° (circles).

Interestingly, the incentive toward wall alignment with the field depends only weakly on the pore width and water-wall contact angle. The net effect therefore originates almost exclusively in the first solvation layers, regardless of the amount of intervening liquid between them. Insensitivity to the nature of wall material, on the other hand, confirms that the observed changes in σ derive primarily from the properties of the liquid, namely its polarity (favoring parallel dipole/wall alignment)[43, 44], as well as orientation dependence of hydrogen bonding[13-15]. Inclusion of specific, angle dependent wall/water interactions[45], present in most hydrophilic materials, would clearly alter and possibly strengthen the dependence of wetting surface free energies on the field angle $\phi$.

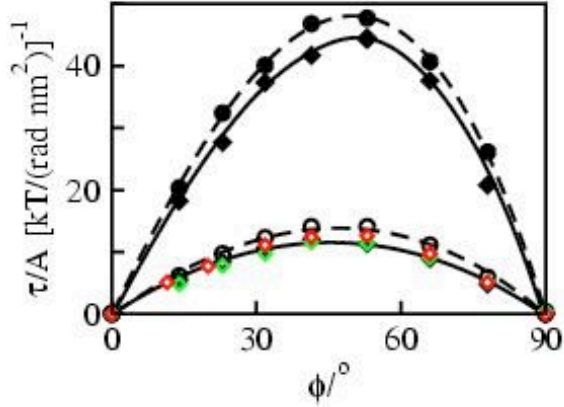

**Fig. 4** Calculated torques on confined water-slab systems subject to applied fields $E_0 = 0.2$ VÅ$^{-1}$ (open symbols) or 0.4 VÅ$^{-1}$ (solid symbols). Diamonds correspond to walls with contact angle $\theta_c = 135°$ and circles are for $\theta_c = 69°$, all for open ($\mu VT$) systems with wall separation $D = 16.4$ Å. The green diamonds describe a closed ($NVT$) system with number of molecules the same as in the field-free systems, while the red diamonds describe a wider pore ($D = 27$ Å), both with hydrophobic walls ($\theta_c = 135°$).

According to Eq. (3), angle dependence of wetting free energy leads to a torque acting on confinement walls. In Figs. 4 and 5, we present the torque dependence on angle $\phi$ measuring the deviation from surface alignment with the field. The curves shown here were obtained by direct calculation of torques on confined water molecules (Eq. 5). They are consistent with results of the alternative calculation, Eq. 3, based on the slopes of wetting free energies presented in Fig. 3. The torque vanishes in free energy extrema corresponding to full alignment ($\phi = 0$) or perpendicular orientation ($\phi = 90°$). In strong fields, the maximal torque (steepest slope of $\sigma(\phi)$) is observed around $\phi \sim 53°$, the limiting angle for a water dipole to accommodate up to three hydrogen bonds[13, 15]. This is different from a symmetric dipolar fluid such as Stockmayer fluid where we find (results not shown) the maximal torque at the expected angle $\phi \sim 45°$.

In agreement with observed variations of wetting free energies, Fig. 3, the torques are virtually independent of confinement width and wall contact angle. Physical reasons for this independence have been discussed in the context of wetting free energies (Fig. 3). In addition, we observe very small effect of ensemble conditions ($NVT$ vs $\mu VT$) on calculated torques acting on the walls.

Unlike calculations of surface wetting free energies, amenable only for the confinement as a whole, direct torque calculations allow us to separate forces acting on individual walls.

Since any significant torque contributions come from water mediation of field forces in surface layers next to either of the walls, net forces on either of the plates can be obtained as sums of molecular contributions allocated to the two halves of the slab divided by the pore mid-plane. Results of such a calculation are presented in Fig. 5. The net torque comprises two unequal contributions from separate plates. In the parallel alignment, the vanishing total torque consists of two finite and oppositely equal torques on the two walls. They are explained by the fact that the spontaneous water alignment differs slightly from exact parallel one[14, 31], causing weak, oppositely equal torques on the two walls. The observed asymmetry in curves $\tau(\phi)$ is directly related to the asymmetry of water molecules and their angle preferences imposed by hydrogen bonding[13-15].

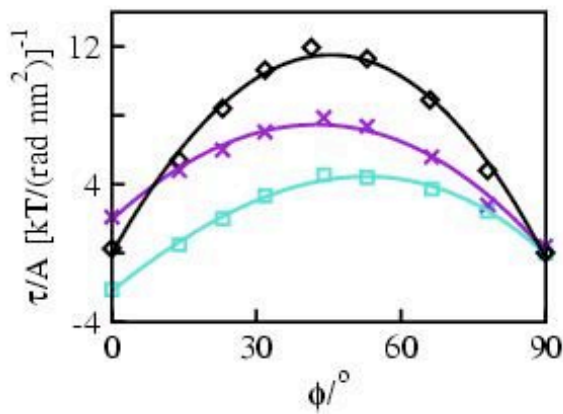

**Fig. 5** Total torques on a confined, 16.4 Å wide water-slab system subject to applied field $E_0 = 0.2$ VÅ$^{-1}$. (black/diamonds), equal to the sum of torques acting on the left (violet/×'s) and right (turquoise/squares) hand side surfaces. For nonzero angles $\phi$, the perpendicular component of the field is pointing towards the right side. The wall-water contact angle is 135°.

Fig. 4 shows a strong dependence of the torque on the field strength. Doubling the field strength brings the value of calculated torques per area from ~ 10 kT rad$^{-1}$ nm$^{-2}$ at $E_0 = 0.2$ V/Å to ~ 40 k$_B$T rad$^{-1}$ nm$^{-2}$, confirming our prediction that $\tau/A$ should be proportional to $E_0^2$. This result, verified in numerous additional calculations (not shown), is consistent with weak-field predictions of polarization energy of water from the Langevin equation, and this is the energy that competes with hydrogen bonding. Deviations from these simple predictions could be expected for more realistic interface models including orientation dependent wall/water interactions[45]. Increasing isotropic wall attractiveness alone gives slightly higher values of torques at hydrophilic walls compared to hydrophobic ones. This is because stronger attraction to the surfaces enhances the impact of orientational preferences of water molecules in the wall/water contact layer compared to more loose contacts next to hydrophobic plates.

So far, our work has only addressed effects unrelated to specific properties of the surfaces. The two general mechanisms contributing to observed torques are based on 1) hydrogen-bond related orientational preferences of interfacial water molecules that already exist in the absence of the field, and 2) the universal advantage of aligning dipoles of any kind *along*, rather than normal to the interface. The general propensity to parallel alignment, supporting favorable head to tail stacking of dipoles, has been analyzed thoroughly in contexts of interfacial structural ordering[43, 44] and anisotropic dielectric response at interfaces[46]. Our calculations using Stockmayer fluid in a field-exposed confinement reveal the existence of similar torques as observed in water, however, these results (not shown) were always insensitive to the sign (polarity) of the field, a phenomenon attributed to asymmetric interactions of surface water molecules.

Lastly we mention the roles of particular water models and boundary conditions we use. We have tested the field response of TIP4P[47] water, and found it to be only marginally more affected by the field as compared to SPC/E, suggesting that our results are not overly sensitive to the water model used. Further, our studies avoid field strengths capable of any significant polarization of water molecules that would warrant the use of polarizable models of water[29, 48]. To assess the impact of truncation of molecular interactions, we have also carried out a number of simulations using the adaptation of 3D-Ewald summation for a slab finite in one of the three spatial dimensions[49]. These additional calculations confirmed (results not shown) the same qualitative behavior, with amplified dependence on field direction observed upon removing the pair potential cutoff we used in most calculations. We also note that in our studies of confinement effects we employ lateral periodicity primarily as a mathematical simplification rather than to describe a truly infinite system. As such, the use of truncated potentials has been argued to produce representative results for nano-sized aqueous confinements encountered in biosystems and in nanomaterials. [50]

## 5 Concluding Remarks

In this work we discuss a new mechanism to orient nanoparticles by an applied electric field when the particles carry no charges or dipoles of their own. Coupling to the field can be accomplished through solvent-mediated interaction between the applied electric field and a nanoparticle. The key to this phenomenon is the orientational preference of solvent molecules in the surface layer. This preference manifests in angle-dependent water/particle potentials that enable interfacial solvent to mediate field-induced orientational forces between dipolar molecules and the nanoparticle. In the case of water, orientation-dependent molecule-wall interactions combine the general propensity of dipolar fluids to spontaneously polarize in a lateral direction[43] with specific effects of water asymmetry and hydrogen bonding[13, 15]. The former effect alone can induce the preference for dipolar alignment with confinement walls and the latter is responsible for polarity dependence (*Janus* interface)[9]. These mechanisms are not limited to confinements, but can equally operate on isolated surfaces.

To maintain the alignment of the nanopore within a tolerance $\delta\phi$, work against the field, $\int_0^{\delta\phi}\tau(E,\phi)d\phi$ must appreciably exceed thermal energy, $k_BT$. For field strength $E_0$ of 0.2 V$\text{Å}^{-1}$ and for a nanopore area of 1 nm$^2$, the simulated torques suggest a typical angle deviation $\delta\phi \sim \pm 20^0$. For a field $E\varepsilon_r$ next to a DNA polyion of $O(1)$ V$\text{Å}^{-1}$, for example, this tolerance diminishes by an order of magnitude, $\delta\phi \sim \pm 2^0$, and the nanoparticles would be fully aligned. In practice, the accessible field strengths, $E_0$, will ultimately determine if the suggested mechanism can one day find use in nanomechanical devices.

We can envisage increasing the sensitivity to the field by using solvents with greater molecular asymmetry and/or dipole moments than water (e. g. DMSO, DMF). Orientational forces are likely to discriminate not only between filled and empty nanopores but also between pores containing solvents of different polarities. While the bigger solvent asymmetry only controls the magnitude of the *Janus* interface effect[9], the solvent polarity combined with molecular asymmetry controls the magnitude of the torque. We plan to explore the use of alternative solvents in our future studies.

**Acknowledgments**

This work was supported by the National Science Foundation through awards CHE-0718724 (to A.L.) and CBET-0432625 (D.B.).